\documentclass[12pt]{article}
\usepackage[left]{lineno}
\usepackage{blindtext}
\usepackage{enumitem}
\usepackage{setspace}
\usepackage{natbib}
\usepackage{graphicx}
\usepackage{epstopdf}
\usepackage[margin=2cm]{geometry}
\usepackage{hyperref}
\usepackage{setspace}
\usepackage{xcolor}
\usepackage{subfig}
\usepackage[us,12hr]{datetime}
\usepackage[version=3]{mhchem}
\usepackage{hyperref}
\usepackage{caption}
\usepackage{array}
\newcolumntype{P}[1]{>{\centering\arraybackslash}p{#1}}
\newcommand{\footremember}[2]{%
   \footnote{#2}
    \newcounter{#1}
    \setcounter{#1}{\value{footnote}}%
}
\newcommand{\footrecall}[1]{%
    \footnotemark[\value{#1}]%
}

\setcitestyle{super,open={},close={}}

\begin{document}
\title{\Large{\bf{A MEMS gravimeter with multi-axis gravitational sensitivity}}}
\author{
    \footnotesize{Richard P. Middlemiss (richard.middlemiss@glasgow.ac.uk),}\footremember{Physics}{Institute for Gravitational Research, S.U.P.A., School of Physics and Astronomy, University of Glasgow, Kelvin Building, University Avenue, Glasgow, G12 8QQ, U.K.} \footremember{Engineering}{James Watt School of Engineering, University of Glasgow, Rankine Building, Oakfield Avenue, Glasgow, G12 8LT, U.K.}
\and \small{\hspace{-0.6cm}Paul Campsie (paul.campsie@strath.ac.uk),}\footrecall{Physics}
\and \small{\hspace{-0.6cm}William Cunningham (william.cunningham@glasgow.ac.uk),}\footrecall{Physics}
\and \small{\hspace{-0.6cm}Rebecca Douglas( rebecca.douglas@hindles.co.uk),}\footrecall{Physics}
\and \small{\hspace{-0.6cm}Victoria McIvor (victoriamcivor@setventures.com),}\footrecall{Physics}
\and \small{\hspace{-0.6cm}James Hough (james.hough@glasgow.ac.uk),}\footrecall{Physics}
\and \small{\hspace{-0.6cm}Sheila Rowan (sheila.rowan@glasgow.ac.uk),}\footrecall{Physics}
\and \small{\hspace{-0.6cm}Douglas J. Paul (douglas.paul@glasgow.ac.uk),}\footrecall{Engineering}
\and \small{\hspace{-0.6cm}Abhinav Prasad (abhinav.prasad@glasgow.ac.uk),}\footrecall{Physics}
\and \small{\hspace{-0.6cm}and G. D. Hammond (giles.hammond@glasgow.ac.uk)}\footrecall{Physics}}

\date{} 

\maketitle

Corresponding authors: Richard P. Middlemiss (+44(0)141 330 5241), and Giles D. Hammond (+44(0)141 330 2258)

\newpage
\section*{Abstract}
A single-axis Microelectromechanical system gravimeter has recently been developed at the University of Glasgow. The sensitivity and stability of this device was demonstrated by measuring the Earth tides. The success of this device was enabled in part by its extremely low resonant frequency. This low frequency was achieved with a geometric anti-spring design, fabricated using well-established photolithography and dry etch techniques. Analytical models can be used to calculate the results of these non-linear oscillating systems, but the power of finite element analysis has not been fully utilised to explore the parameter space before now. In this article, the results of previous analytical solutions are replicated using finite element models, before applying the same techniques to optimise the design of the gravimeter. These computer models provide the ability to investigate the effect of the fabrication material of the device: anisotropic $<$100$>$ crystalline silicon. This is a parameter that is difficult to investigate analytically, but finite element modelling is demonstrated here to provide accurate predictions of real gravimeter behaviour by taking anisotropy into account. The finite element models are then used to demonstrate the design of a three-axis gravimeter enabling the gravity tensor to be measured - a significantly more powerful surveying tool than the original single-axis device.

\section{Introduction}
\label{intro}
Gravimeters have applications in air and land-based oil and gas exploration\cite{Barnes2012,Rim2015}, sinkhole analysis \cite{Kaufmann2014}, the detection of subterranean tunnels and cavities \cite{Butler1984,Romaides2001}, CO$_2$ sequestration \cite{Gasperikova2008}, geothermal reservoir monitoring \cite{Nishijima2016}, archaeology \cite{Panisova2009}, hydrology \cite{Jin2013,Fores2017}, and volcanology \cite{Fernandez2017,Carbone2017,Aparicio2014,Battaglia2008,Rymer2000a}. Commercial gravimeters are all expensive ($\>\$$100,000 USD) and use a range of different technologies for different applications. The commercially available gravimeters all require levelling, and this is carried out manually or be incorporating additional components to automate this process. 

In previous work\,\cite{Middlemiss2016,Campsie2016,Middlemiss2016a}, the development of a low frequency microelectromechanical system (MEMS) gravimeter with a sensitivity of 4$\times10^{-7}$\,ms$^{-2}/\sqrt{\mathrm{Hz}}$ was discussed. This device has since been miniaturised and undergone field-testing\,\cite{Middlemiss2017,Bramsiepe2018,Middlemiss2018}. A series of sensors are currently being built for integration within the NEWTON-g volcano gravity imager at Mt Etna, Sicily \cite{Carbone2020}. The device is capable of high acceleration sensitivity in part because of its extremely low resonant frequency MEMS resonator. This resonant frequency was achieved via the use of a geometric anti-spring design for the mass-on-spring system. A low resonant frequency means that the ratio is minimised between the spring constant, $k$, and the mass of the proof mass, $m$, giving a larger displacement for a given acceleration, and thus greater potential sensitivity for the gravimeter.

An anti-spring can be characterised as having a negative or at least partially negative restoring force. As an anti-spring is extended, the spring constant decreases. One way to design an anti-spring is by using curved monolithic cantilevers, connected at a central point to constrain the motion vertically\,\cite{Ibrahim2008}. Monolithic geometric anti-springs are used as low frequency seismic isolators in the VIRGO gravitational wave detector\,\cite{Bertolini1999,Cella2005,Acernese2015a,Abbott2016}. With the aim of achieving a high acceleration sensitivity for the MEMS gravimeter a monolithic geometric anti-spring configuration was chosen. A monolithic geometry was important because this allowed for the device to be fabricated from a single silicon chip.

An elegant analytical description of monolithic geometric anti-springs (as used in the VIRGO detector) was written by Cella et. al.\,\cite{Cella2005}. In this mathematical solution, the authors devise simplified models to describe the behaviour of this non-linear spring -- taking a Lagrangian approach. They do so in order to avoid `the brute force of finite element analysis'. It is our belief, however, that there is also value in conducting finite element (FE) analysis on anti-spring systems. It allows one to investigate the physical consequences of making very small adjustments to the design. Furthermore, with the advance in computing power since 2005 -- when Cella et. al. published -- one has the option to be more carefree with computer models that require significant processing power. Little work has been conducted in the FE analsysis of geometric anti-springs since this time. One exception is the recent work of Zhang et. al. \cite{Zhang2020}, who have explored the use of FE analysis to assess the thermal behaviour of geometric anti-springs. Zhang et. al., however, did not demonstrate that the computational results were self-consistent with earlier analytical works in the literature. 

Here, the results of Cella et. al.\,\cite{Cella2005} are replicated, first analytically, and then in an FE model. After demonstrating that these two models are in agreement, it is demonstrated that FE modelling offers great advantages to the design process. The crystal structure of the fabrication material -- $<$100$>$ silicon -- can be included in the FE model. Since this material has an anisotropic Young's Modulus, modelling such a parameter in an analytical model would be extremely difficult. Finally, it is demonstrated that by tuning the MEMS resonators in the FE model it is possible to construct a system of three identical devices in a triaxial configuration\,\cite{Galperin1955} that would have acceleration sensitivity in $x$, $y$, and $z$. A system that enables the measurement of the vector components of gravity would not need the same stringent levelling before measurements are undertaken. This would improve the practical use of the gravimeter for a range of applications.

\section{Results}
\subsection{Analytical Model}
\label{sec:analytical}

\begin{figure}
\includegraphics[width=\columnwidth]{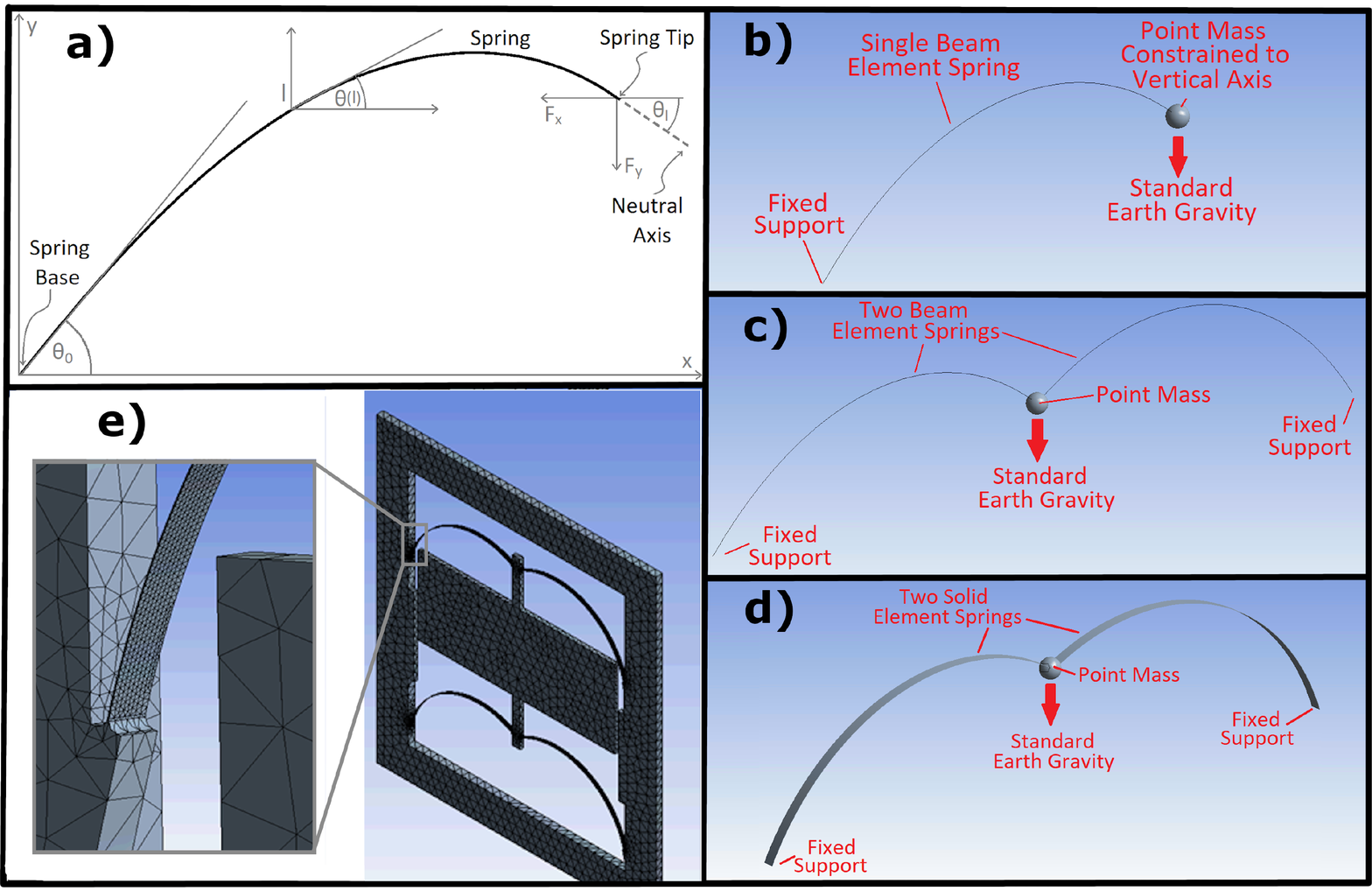}
\caption{\label{fig:geom}The geometries of the springs used in the ANSYS finite element models. {\bf{a)}} A simplified 1-dimensional analytical schematic diagram of the geometric anti-spring. {\bf{b)}} The single spring model with boundary conditions applied to the free end to fix the displacement and rotation. A point mass of $8\times10^{-6}$ kg is applied to the free end of the spring. This system is modelled using beam elements. {\bf{c)}} The same model as {\bf{a)}}, but with two symmetric springs and a point mass of $1.6\times10^{-5}$ kg. In this model the boundary conditions on the free end are removed. {\bf{d)}} The same design as \ref{fig:geom}c but with solid elements.  In all cases a load step of 1 is equal to applying an acceleration of 9.8066 m/s$^2$. {\bf{e)}} The final design of the 4-spring geometric anti-spring MEMS sensor, as modelled in ANSYS. The results of models conducted with this geometry would only converge if a dense mesh density was utilised over the springs.}
\end{figure}

In order to verify the FE models of the MEMS based geometric anti-spring, a simple analytical model -- based on the work of Cella et.al.\,\cite{Cella2005} -- was created. This model is displayed in figure \ref{fig:geom}a. The model outlines the parameters of a single geometric anti-spring blade. Such blades are never used individually; the proof mass is suspended from two or more symmetric blades, constraining it to move along a vertical axis; thus reducing the analytical problem to one dimension. The spring is clamped at the base with a launch angle of $\theta_0$ and is constrained at the proof mass, or spring tip, by an angle $\theta_L$. This results in a boundary value problem that can be conveniently solved in MATLAB with the \emph{bvp4c} algorithm\,\cite{Shampine2000}. In order to achieve this solution the system must be broken down into a pair of coupled first order equations, which can be defined in terms of the curvilinear coordinate along the blade, $l$\, \cite{Cella2005}.

\begin{equation}
\frac{dp}{d\epsilon} = G_x\sin{\theta}-G_y\cos{\theta}
\label{eq:first_order_1}
\end{equation}

\begin{equation}
\frac{d\theta}{d\epsilon} = \gamma(\epsilon)p
\label{eq:first_order_2}
\end{equation}

\noindent{where $\theta$ is the angle that the spring arc makes with the neutral axis at a given position $(x,y)$; $\epsilon = l/L$ (where $L$ is the total arc length of the spring); $\gamma(\epsilon) = w(0)/w(L)$ is the spring's profile where $w()$ is the spring width at the base $(0)$ and at the end $(L)$ respectively. The `width' of the springs correspond to the dimension into the page in figure \ref{fig:geom}a. The springs utilised in gravitational wave seismic isolation systems are triangular in order to maintain a constant stress along their length i.e. their width varies along the length of the spring. Due to the limitations of etching silicon, however, the MEMS springs are rectangular and thus $\gamma(\epsilon) = 1$ for this work. The parameters $G_x$ and $G_y$ are dimensionless forces that are applied to the spring tip. These are related to the real forces via:

\begin{equation}
G_i = \frac{12L^2}{Ed^3w(0)}F_i \;\;\;\;\;\;\; \text{with}\;i=x,y
\label{eq:dimensionless}
\end{equation}

\noindent{where $E$ is the Young's modulus, and $d$ is the thickness of the spring. Finally, the $(x,y)$ coordinates of the spring profile are determined from:

\begin{equation}
x = L\int_0^1\sin({\theta}(\epsilon))d\epsilon
\label{eq:x}
\end{equation}

\begin{equation}
y = L\int_0^1\cos({\theta}(\epsilon))d\epsilon
\label{eq:y}
\end{equation}

As mentioned above, the tip of the geometric anti-spring is constrained to move along a vertical line. This is equivalent to requiring that the horizontal position of the tip always maintains a constant value. This constraint is defined via the compression ratio $x_{\text{com}}=x_{\text{tip}}/L$. As the vertical displacement and/or compression ratio increases, the $2^{\text{nd}}$ spring provides the force $G_x$, which introduces the negative component of the spring constant (the $2^{\text{nd}}$ spring is not displayed in figure \ref{fig:geom}a, but it is symmetrically mirrored in the vertical plane that intersects the spring tip). This is the reason why the term ``geometric anti-spring" is used; the anti-spring nature comes from geometry alone. As demonstrated in Cella\,\cite{Cella2005}, for certain compression ratios the spring exhibits a bifurcation in its force-displacement curve resulting in two stable operating points. This is not a desirable situation, and so in this paper only solutions that exhibit a single operating point are considered. In a displacement-force graph this will be the position of maximum gradient.

The MEMS devices were fabricated with springs with zero initial tension and the loading is provided via gravity alone. A different approach is therefore taken to solving the equations described in Cella\cite{Cella2005}; at the initial starting point the forces must be $G_x=G_y=0$. For the simple analytical model, the following values were used: $E=169$ GPa, $d=7$ $\mu$m, $w=240$ $\mu$m and $L=5.48$ mm. This matches the typical MEMS gravimeters previously fabricated\cite{Middlemiss2016}. The FE solutions of such systems will be described later in section \ref{sec:Full_FEA}. A launch angle of  $\theta_0=\pi/4$ was utilised, along with a final angle $\theta_L=\pi/6$. For zero initial starting force and an arc length of $6$ mm the resulting compression ratio is 0.933, which as demonstrated in Cella produces a stable system. For the MEMS system, operating at smaller compression ratios can be achieved simply by increasing the launch angle such that the arc length increases while the tip remains at the same position. The solution for zero initial force $(G_x=G_y=0)$ results in a spring which follows a simple arc. For an arc length of $6$ mm, $\theta_0=\pi/4$ and $\theta_L=\pi/6$ the radius of the arc is $4.58$ mm.

The procedure for analytically generating the force-displacement graph is as follows. A vertical load component is selected and the boundary value problem solved. The value of the horizontal force is then adjusted until the desired spring tip horizontal compression is achieved. For these models a tolerance was chosen of $1\times10^{-6}$. The value of the vertical force and vertical spring tip were then stored, and the process repeated for different values of vertical force. Figure \ref{fig:max_grad} displays the results of the analytical model together with those of the FE model, which will be described in the following section. In this figure the y-axis is the calculated vertical spring tip displacement, and the x-axis is the equivalent load step. In this case a load step of 1 is equivalent to a force of $7.8\times10^{-5}$ N (since a point mass of $8\times10^{-6}$ kg corresponds to the mass of previously fabricated MEMS devices)\cite{Middlemiss2016}.

\begin{figure}
\includegraphics[width=\columnwidth]{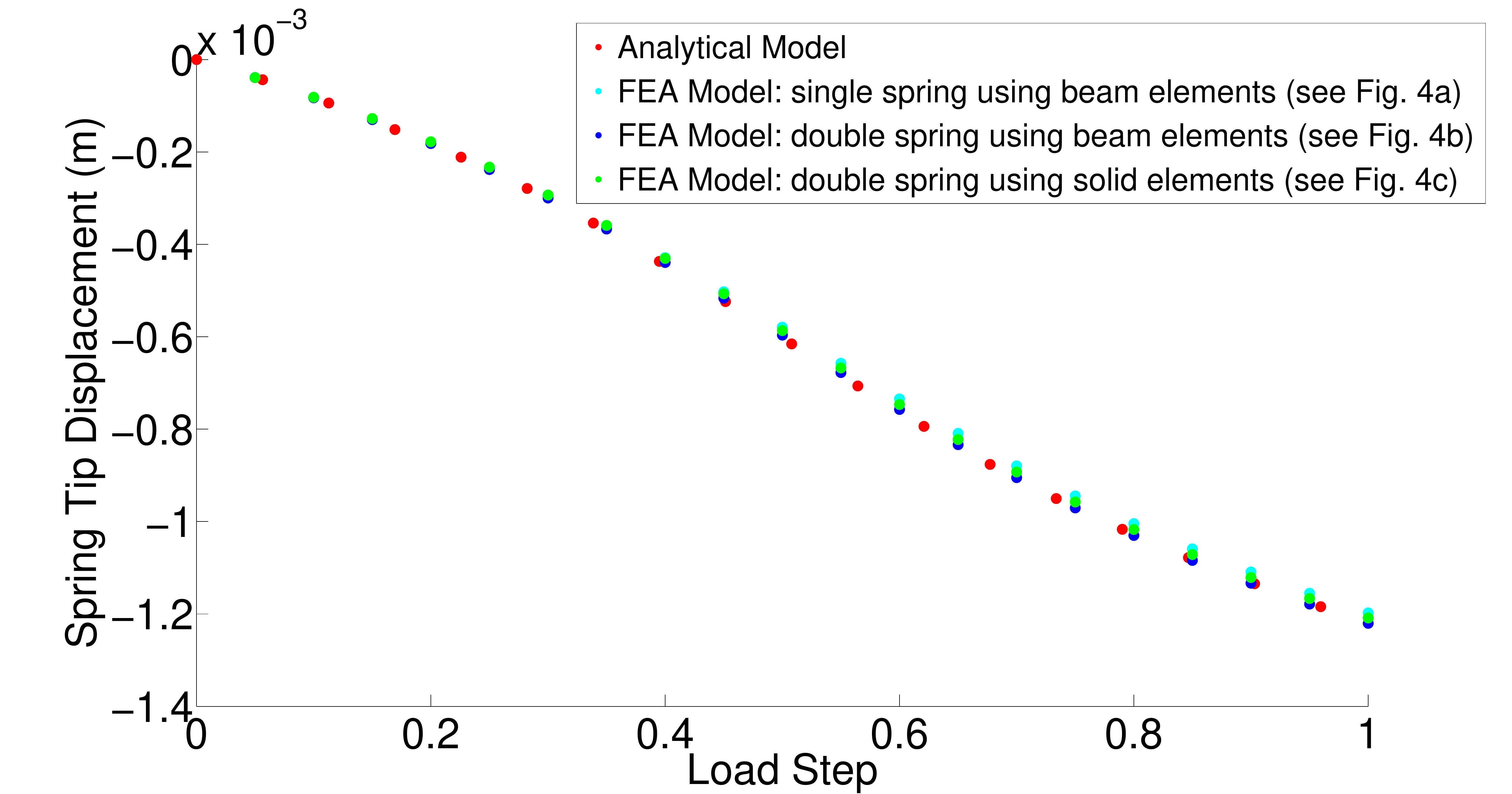}
\caption{\label{fig:max_grad}FE Vs. Analytical Force-Displacement Comparison. A displacement-force graph comparing the MATLAB analytical model (red series) and the ANSYS finite element models. The cyan series is the FE model for a single spring, modelled using ANSYS beam elements. The blue series is the FE model for a double spring, again using beam elements. The green series is also a double spring configuration, but ANSYS solid elements have been used instead. The position of maximum gradient corresponds to the stable oscillation point. }
\end{figure}

The reduction in the gradient of this graph is equivalent to the system softening around a working point of $0.8$ mm vertical displacement. Either side of this point the system is stiffer, therefore the system is a stable resonator at this working point. This is easily observed by calculating the resonant frequency, $f$, of the system using equation \ref{eq:frequency}:

\begin{equation}
f = \frac{1}{2\pi}\sqrt{\frac{k}{m}}
\label{eq:frequency}
\end{equation}

\noindent{where $k$ is the spring constant and $m$ is the mass suspended from the spring. $k$ can be determined by numerical differentiation of the force-displacement curve in figure \ref{fig:max_grad}. The results of this differentiation are displayed in the red series of figure \ref{fig:freq} (along with the results of the corresponding FE models that are discussed in section \ref{sec:FEA}). A stable operating point is demonstrated at a load step of around 0.55, where it can be observed the data goes through a turning point.

\begin{figure}
\includegraphics[width=\columnwidth]{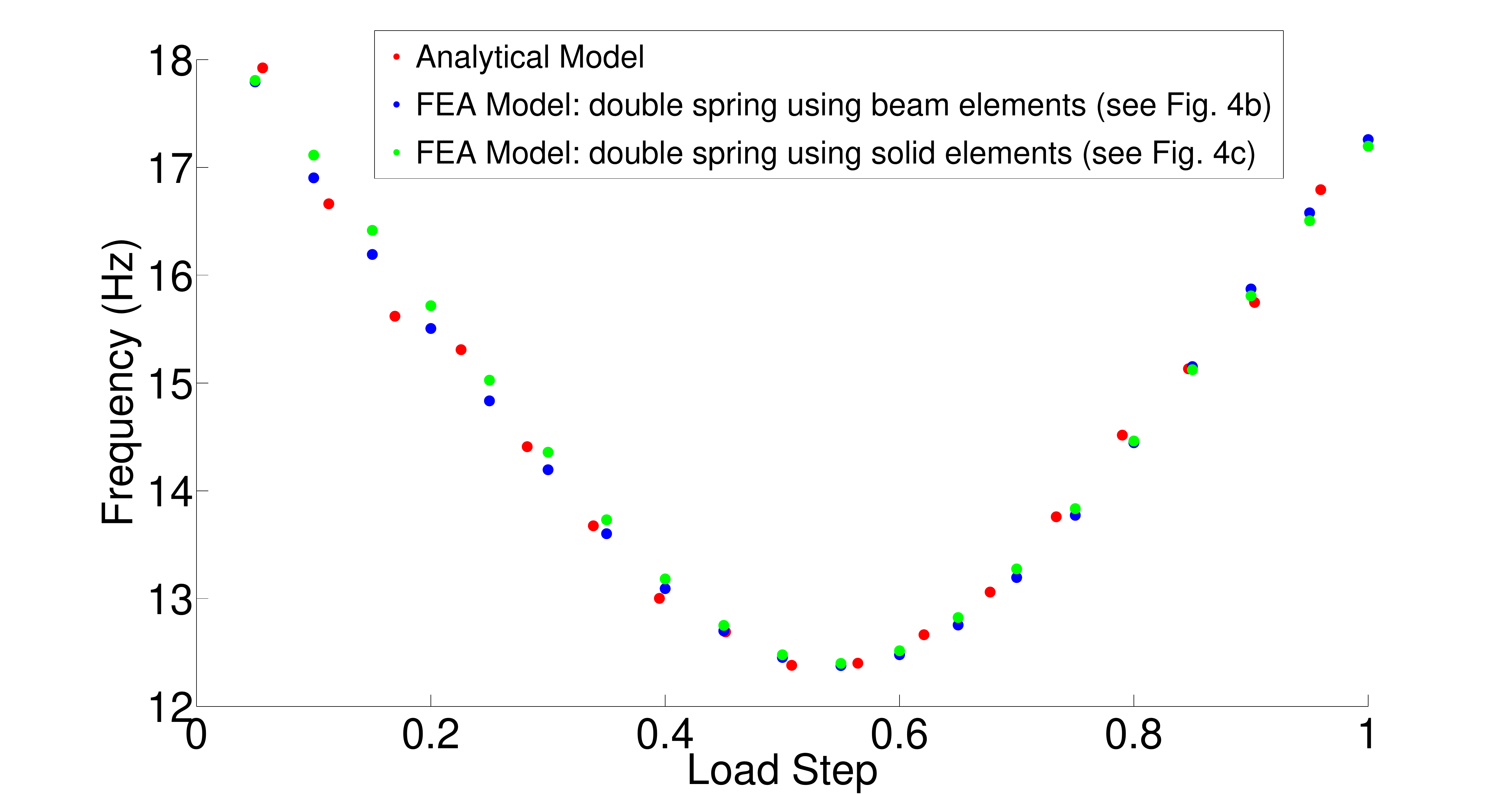}
\caption{\label{fig:freq}FE Vs. Analytical Frequency Comparison. A comparison of the resonant frequencies of the MATLAB analytical model (red) and the ANSYS finite element models. The blue series is the FE model for a double spring, modelled using ANSYS beam elements. The green series is a double spring configuration, modelled using ANSYS solid elements. It can be observed that as the loading on the spring increases, the resonant frequency decreases, goes through a minimum, and then increases again. }
\end{figure}

\subsection{Finite Element Model}
\subsubsection{Initial Model}
\label{sec:FEA}

In order to test the validity of the analytical model a series of simple spring systems were built using ANSYS workbench v17 software. The computer aided drawing (CAD) package within the software -- ANSYS Design Modeller -- was used to generate these systems. The geometric parameters (launch angle etc.) of the springs in each model were identical to those used in the analytical system described in section \ref{sec:analytical}. Three initial FE simulations were performed to verify that the ANSYS model could correctly determine the displacement-force curves of the geometric anti-spring system. The use of both beam elements and solid elements were tested. These models are displayed in subfigures \ref{fig:geom}b,c and d).

In the first instance a single spring was designed (see figure \ref{fig:geom}B) and a point mass applied to the free end. The tip of the spring had boundary conditions on the free end that ensured that there was no rotation or horizontal displacement. These boundary conditions were needed to model the constrained geometric anti-spring. A fixed clamp was applied to the base and a point mass of $8\times10^{-6}$ kg attached at the spring tip. A simple vertical acceleration was then applied to the system with maximum value equal to $9.8066$ m/s$^2$ to simulate a gravitational loading. A force-displacement plot was generated using the results of this simple model. This data is displayed in the light blue series of figure \ref{fig:max_grad}.

The second model to be analysed was a more realistic test: a full two-spring system. The system was built using both beam elements (figure \ref{fig:geom}c) and solid elements (figure \ref{fig:geom}d). The use of beam elements allows the model to converge quickly, and with 1 $\mu$m wide beams the model only takes 30 s to run. The solid elements are much more computationally intensive, requiring several minutes to execute. The general approach in modelling these systems is to fix the geometry with the analytical MATLAB model and only utilise the full FE simulation to extract stress, investigate etching tolerances, and to investigate the behaviour of crystalline silicon. Since this system included two springs, the boundary condition used to fix the horizontal position of the tip and the spring rotation was removed. A single point mass of $1.6\times10^{-5}$ kg  was applied to the centre point of the springs and the system was allowed to evolve with a ramped acceleration. For all models a uniform load step was maintained for the solver and split into 20 separate iterations. The vertical force was determined from the value of the acceleration and the point mass, while the position of the vertex of the spring was used to determine the vertical position. As expected, the horizontal position of the spring remained unchanged due to the symmetry of the system. This data for both of these models is displayed in figure \ref{fig:max_grad}. The dark blue series represents the beam element model, and the green series represents the solid element model.

The force-displacement data for the two double spring models were used in turn to calculate the resonant frequency as a function of load step. This was carried out via numerical differentiation. This data is displayed in figure \ref{fig:freq}. Alongside the analytical data (red series), the beam element model is represented by the dark blue series, and the solid element model is represented by the green series.

It can be observed from figures \ref{fig:max_grad} and \ref{fig:freq} that there is excellent agreement between the analytical and FE models.  This agreement is evidence that the simple analytical model can be utilised to test and optimise new geometries. Conversely, the full FE model is useful for determining the effect of varying spring geometries caused by non-ideal etching tolerances; in addition to exploring the stress in the springs and the effect of crystalline silicon (and its non-axisymmetric Young's modulus).

\subsubsection{Crystalline Silicon Elastic Modulus}
Silicon is a crystalline material and thus exhibits a Young's modulus that depends on the orientation of the crystal axis to the etched device. It is particularly important to utilise the correct Young's modulus in FE simulations in order to accurately predict the ultimate displacement/resonant frequency of the MEMS device since the moduli can vary by up to 45$\%$\,\cite{Hopcroft2010} depending on the axis. Furthermore, FE simulations also have the benefit of fully modelling the shear/strain properties of a spring, although the effect of shear in thin springs (such as those of the MEMS device) is negligible. The MEMS devices are fabricated from the $<$100$>$ plane of a silicon wafer, which provides the lowest Young's modulus of any of the possible orientations. The modulus tensor provided by Hopcroft et. al.\,\cite{Hopcroft2010} was utilised, which is given by:

\begin{equation}
\begin{bmatrix}
\centering
   \bf{194.5} & \bf{35.7}  & \bf{64.1}  & \bf{0}    & \bf{0}    & \bf{0}   \\
    \bf{35.7}  & \bf{194.5} & \bf{64.1}  & \bf{0}    & \bf{0}    & \bf{0}    \\
    \bf{64.1}  & \bf{64.1}  & \bf{165.7} & \bf{0}    & \bf{0}    & \bf{0}   \\
    \bf{0}     & \bf{0}     & \bf{0}     & \bf{50.9} & \bf{0}    & \bf{0}    \\
    \bf{0}     & \bf{0}     & \bf{0}     & \bf{0}    & \bf{79.6} & \bf{0}    \\
    \bf{0}     & \bf{0}     & \bf{0}     & \bf{0}    & \bf{0}    & \bf{79.6} \\
\end{bmatrix}
\label{eq:Si_matrix}
\end{equation}

\noindent{where each element has units of GPa. It is important to note that ANSYS differs in its standard definition of the stifness tensor compared to that of Hopcroft. This can be remedied by switching elements 44 and 66.

\subsubsection{Tuning the System Behaviour}
\label{sec:tuning}
For gravimetry applications there is a desire to develop low frequency resonators that provide stable behaviour. It would therefore be beneficial to have a simple means of tuning the resonant frequency for a fixed proof mass. This is generally achieved by reducing the compression ratio of the spring system $x_{\text{com}}=x_{\text{tip}}/L$. For MEMS systems, however, in which the springs are etched, there is no capability to actually change the horizontal compression ratio as this is just set by the initial geometry of the MEMS mask. As mentioned earlier, a convenient way to alter the frequency is to change the launch angle of the spring. This has the effect of increasing the arc length of the spring, $L$, which in-turn reduces the compression ratio for a given $x_{\text{tip}}$. Figure \ref{fig:launch} shows the effect on the resonant frequency for a selection of launch angles modelled via ANSYS. A two-spring system with a mass of 1.6$\times10^{-5}$ kg was utilised with a maximum acceleration of 9.8066 m/s$^2$. As the launch angle increases the resonant frequencies drop. Although the 55$^{\circ}$ launch angle does not reach a minimum by load step 1, running this model with a larger point mass results in a minimum frequency of 5 Hz.

\begin{figure}
\includegraphics[width=\columnwidth]{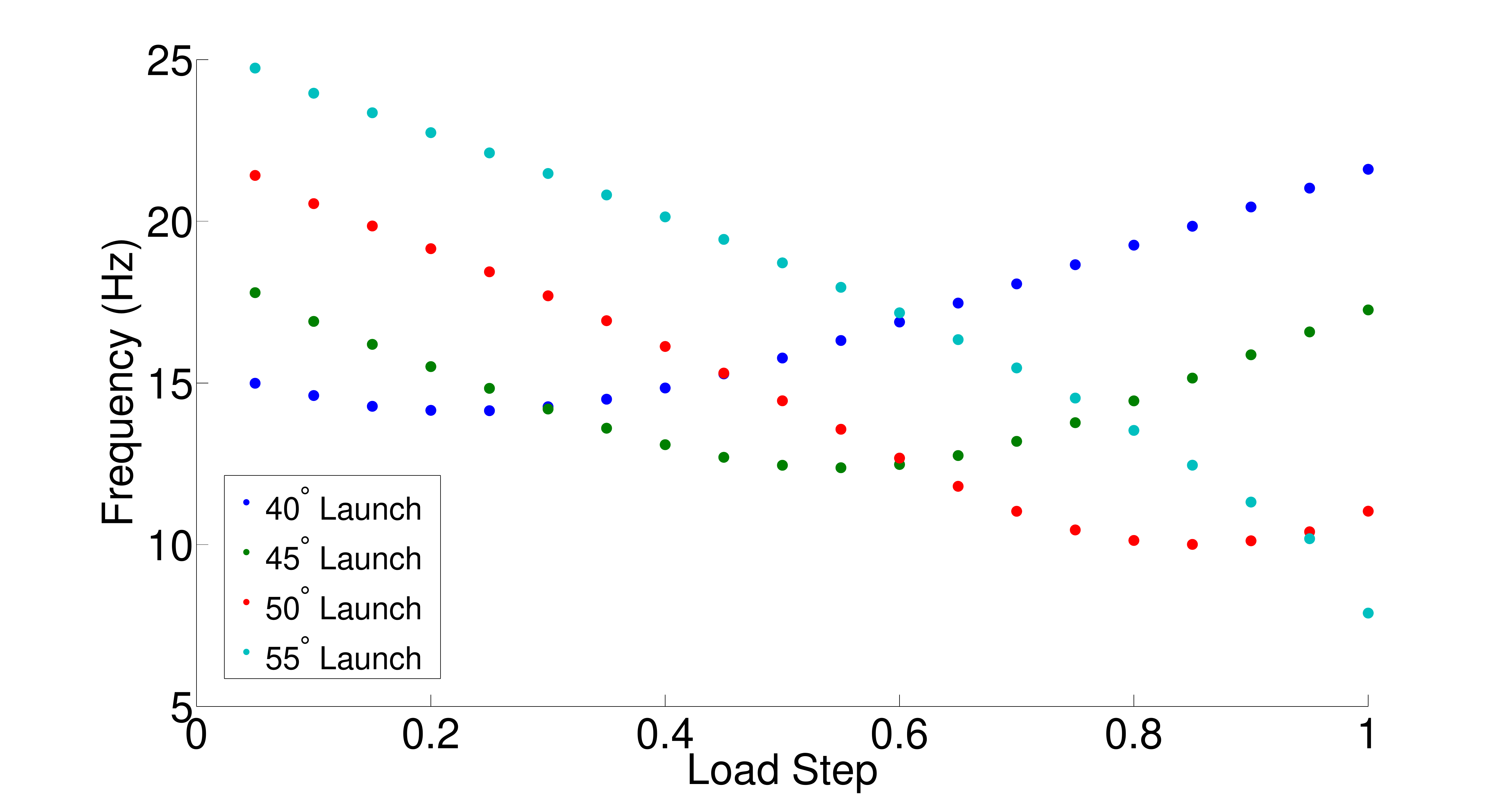}
\caption{\label{fig:launch}Launch Angle Comparison. The resonant frequency of a double spring system for different launch angles (see figure \ref{fig:geom}a). Increasing the arc length, via the launch angle, is a convenient method to provide a lower compression ratio in a MEMS geometry that can only be fabricated with a single fixed horizontal spring extent. The greater the launch angle, the lower the resonant frequency.}
\end{figure}


An additional means of tuning the system is to alter the ratio of $k/m$. By reducing the spring thickness \emph{or} increasing the mass, the level of loading at which the oscillator reaches its minimum frequency can be changed. Figure \ref{fig:launch_min} demonstrates that changing the thickness of the spring (whilst keeping the mass constant) does not change the minimum frequency of the system substantially (compared to the change that can be induced by altering the launch angle). The following protocol could therefore be followed in order to tune the design of a MEMS geometric anti-spring. First, the launch angle could be chosen to determine the minimum frequency. With this launch angle fixed, the thickness of the springs could then be altered in order to set the loading at which this minimum frequency is achieved. For situations in which changes in vertical gravity need to be measured, it would be necessary for the frequency minimum to occur at full loading (i.e. suspended vertically in the Earth's gravitational field). It is not always the case that gravimeters would be operated in such a vertical configuration, as will be discussed in Triaxial MEMS Gravimeter section.

\begin{figure}
\includegraphics[width=\columnwidth]{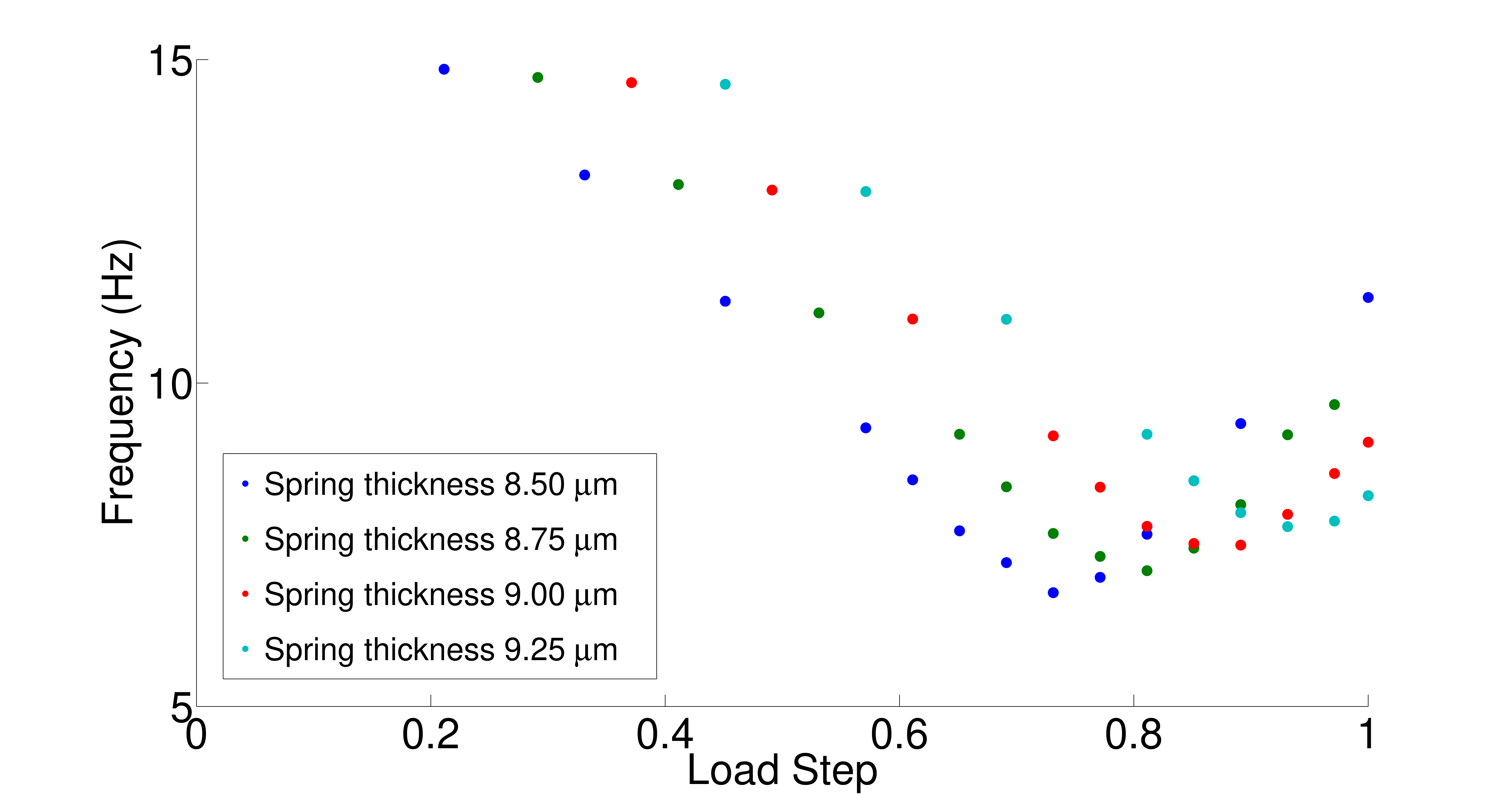}
\caption{\label{fig:launch_min}Spring Thickness Comparison. The resonant frequency of the double spring system for different spring thicknesses (whilst keeping the launch angle constant). The level of loading at which the frequency minimum occurs can be tuned by altering the thickness of the spring.}
\end{figure}

\subsubsection{Full Finite Element model of the 4-spring MEMS system}
\label{sec:Full_FEA}
The next stage of the investigation was to model a full MEMS gravimeter. Such a device requires four springs to support a central mass. With only two springs the system would only be stable when vertical: if rotated sideways torsional stresses would break the springs. Four springs are utilised so that the system is self-supporting in any orientation. Although the simpler devices analysed in this paper were drawn in ANSYS Design Modeller, these more complex designs were drawn in SolidWorks and exported using the STEP2014 format into ANSYS Modeller via the External Geometry Import. This also works well and offers a more convenient interface to model complex geometries and parametrically combine dimensions.

The final design, comprised of four springs, is displayed in figure \ref{fig:geom}e. This design is essentially a double version of the system presented in figure \ref{fig:geom}d. The spring launch angle has been increased to 60$^{\circ}$, which provides a low frequency but stable oscillator. The arc length for this system is 8.33 mm which results in a compression ratio of 0.87. While smaller compression ratios are possible, a compromise was chosen between the lowest resonant frequency and the total stress in the spring. This final design has a proof mass of $3\times10^{-5}$ kg suspended from the four springs. The boundary condition applied to the system is a fixed base on the reverse of the wafer. Again, the acceleration was ramped but rather than requiring a constant load step the ANSYS programme it was allowed to determine the most appropriate steps. Typically a step size of $1\times10^{-2}$ was used at first (where 1 is equal to the full load step of 9.8066 m/s$^2$). As the model converged, the step was typically increased.  Around the area of maximum deflection, however, the model often bisects and the load step is reduced. This bisection occurs because the spring system becomes more non-linear at this stage, due to the effect of the geometric anti-spring.

In this model a minimum element size of 4.25$\times$10$^{-6}$ m was utilised. This system has a minimum resonant frequency when vertical of around 5 Hz and a maximum equivalent stress in the spring of 200 MPa. From our previous studies, and measurements of the breaking stress in in thin silicon suspension beams\,\cite{Cumming2014} this seems an appropriate level of stress to provide a robust device. A very dense mesh is required across the springs of the model, especially where the springs meet the frame and the proof mass. This is where most of the motion and therefore most of the stress is concentrated. Tests were carried out on the mesh density to confirm that the model converges.

\subsection{Experimental Verification of Finite Element Modelling}
\label{sec:Experiment}
To test the efficacy of the finite element modelling, a comparison was made between the results of these computer models and some real devices. Two FE computer models were run, both with a 60$^{\circ}$ launch angle, the same arc length of 8.33 mm, and the same corresponding compression ratio of 0.87. One model, however, utilised the anisotropic Young's Modulus tensor presented in equation \ref{eq:Si_matrix}; and the other model used a more simplistic option: instead of an elasticity that changes with crystal axis, an isotropic value of 169 GPa was used. This value was selected as the best approximation of a global Young's modulus for a device fabricated from $<$100$>$ silicon. For each of these models, the variation of frequency as a function of spring thickness was calculated. The resulting data from the anisotropic model is presented in black in figure \ref{fig:Exp_Results}, and the data from the simplistic isotropic model is presented in red on the same graph.

A series of 7 MEMS devices were fabricated using the design presented in figure \ref{fig:geom}e. Using the same launch angle of 60$^{\circ}$, the same arc length of 8.33 mm, and the same corresponding compression ratio of 0.87; this design was patterned onto the surface of a photoresist-coated, 240 $\mu$m thick piece of silicon wafer using standard photolithography techniques. The silicon was then through-wafer etched using a highly anisotropic deep reactive ion etch (DRIE). Once complete, the MEMS devices were each mounted vertically in a bracket. In turn, each MEMS device was excited using a high impulse tap on the workbench; the resultant oscillation was filmed to ascertain the primary resonant frequency of each device. The films were recorded using a video camera with a 210 frame per second recording rate (the frame rate of the camera was confirmed by filming a stopwatch for a period of 30 s). For each of the 7 videos, 20 oscillations were counted and the number of frames recorded. The corresponding frequencies of each of the 7 MEMS devices was calculated: the samples had a mean frequency of 12.92 Hz, with a standard deviation of 0.36 Hz.

The sample with the closest resonant frequency to the mean of the group (12.88 Hz) was analysed with an optical microscope and measurements were made of the spring thickness. A measurement was made at both ends and the middle of each of the four springs. Measurements were taken on both sides of the spring (top and bottom) to account for any anisotropy in the DRIE etch. A further measurement was made of the profile of a broken spring using a scanning electron microscope (SEM). From this image it was clear that a slight undercut occurred during the etch, meaning that the measurements of the top of the spring were 1.2 $\mu$m larger than the true thickness of of the spring. The SEM indicated that the spring thickness remained constant from top to bottom apart from the final few microns, where it was observed that a small amount of notching \cite{Wu2010} had occurred. This made the optical measurement of the bottom thickness an unreliable indicator of the true spring thickness. It was therefore decided that the best indicator of the spring thickness was the optical measurement of the top of the spring, with a statistical correction made to account for the consistent undercut profile. The mean spring thickness was taken to be 8.86 $\pm$ 0.12 $\mu$m, where the error is given by the standard error of the mean of the population (of 12 measurements). This experimental result is presented as a blue data point in figure \ref{fig:Exp_Results}, with the standard error of the thickness measurement indicated by a horizontal error bar. The fact that the anisotropic silicon FE model lies within the bounds of the experimental result (whereas the isotropic FE model does not) provides validation that taking into account a variable Young's Modulus is important for the prediction of the physical behaviour of a fabricated device. Furthermore, since this anisotropy is not possible to model analytically, this result is also validation that FE modelling provides an improvement in predictive accuracy compared to present analytical models

\begin{figure}
\includegraphics[width=\columnwidth]{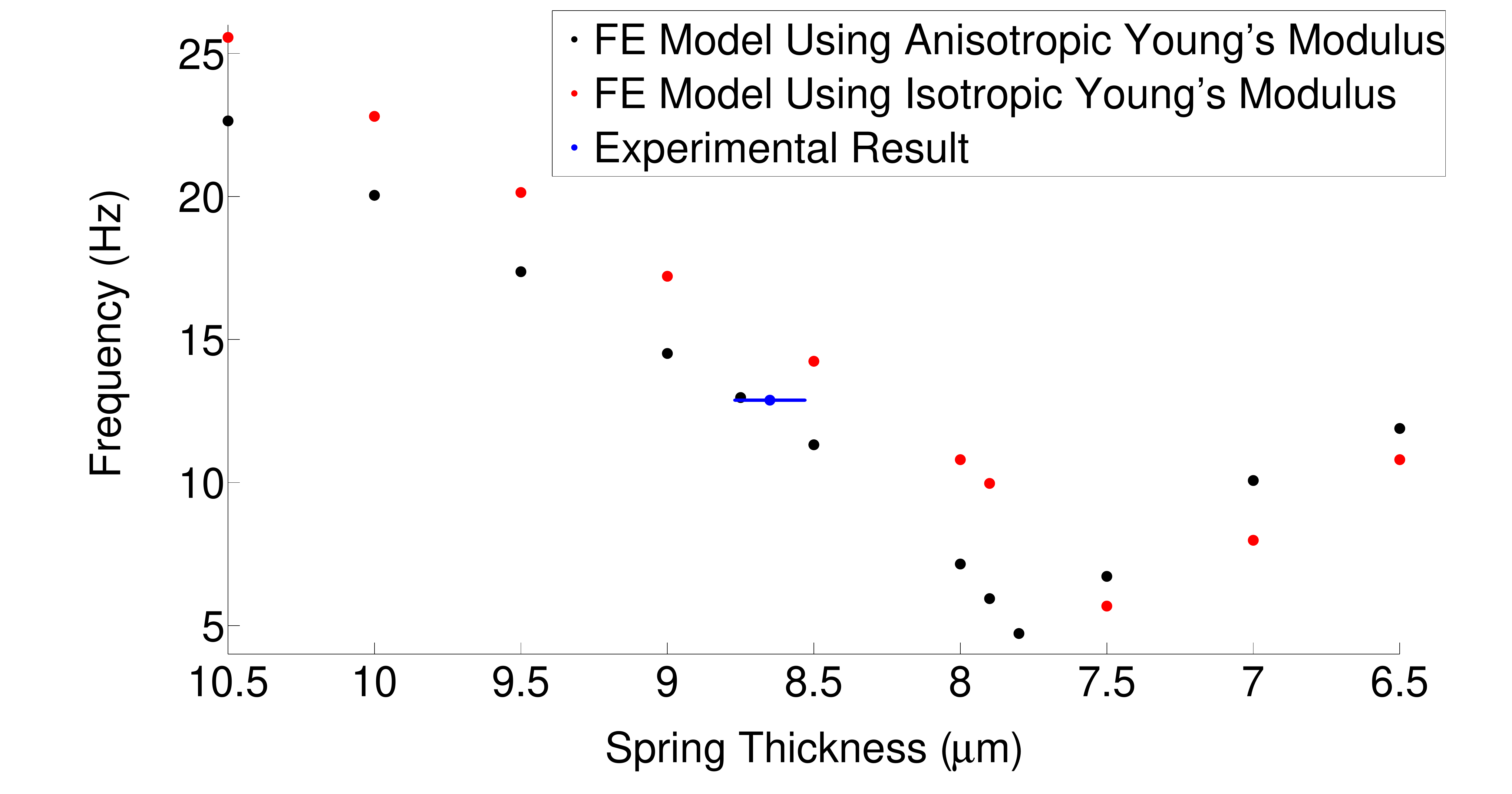}
\caption{\label{fig:Exp_Results}Experimental Comparison to FE Model. The red series is a FE model in which an isotropic Young's Modulus of 169 GPa has been used. The black series shows a the results of a FE model in which the physical anisotropy of silicon has been accounted for: the Young's Modulus is inputted to the model as a tensor (see equation \ref{eq:Si_matrix}) since Young's Modulus varies between crystal axes. The blue data point represents a real sample, fabricated using the same design as that used in the FE models. The error bar on this data point is the standard error of the 12 spring thickness measurements.}
\end{figure}

\section{Discussion}
\subsection*{The Design of Future Triaxial MEMS Gravimeters}
\label{sec:Galperin}

After experimental verification FE model accuracy, more complex designs of gravimeters could be considered. All previous works on this topic by the authors have concentrated on gravimeters that operate in a vertical configuration. These cannot be used to measure variations in acceleration other than those in the vertical, $z$, component of gravity. It is possible, however, to design a gravimeter that has sensitivity to the $x$, $y$, and $z$ components of gravity, not just $z$. Such a gravimeter has the advantage that providing the orientation of the device is known, the levelling requirements are not as stringent as those for a one dimensional gravimeter measuring only the $z$-component. Making such a device is only possible if the angle at which the minimum frequency (where the optimum acceleration sensitivity occurs) is tunable. As has been demonstrated in figures \ref{fig:launch} and \ref{fig:launch_min}, it is possible to tune both the minimum frequency, and the loading at which this frequency occurs. Since the angle of the device is just a proxy for loading (along with spring thickness, and the mass value), one could easily design a device that would reach its minimum frequency at a specific angle. In such a device three tuned MEMS devices could be placed in the Galperin configuration\,\cite{Galperin1955} at an angle of $\theta = 54.7^{\circ}$ from the horizontal, and separated azimuthally from each other at an angle of 120$^{\circ}$ (see Fig. \ref{fig:Galperin}). The Galperin configuration was designed to allow three identical sensors to measure gravity (or seismic activity) in three dimensions. Conversely, if one wanted to mount three sensors parallel to $x$, $y$ and $z$, then two different sensor geometries would be required. This is because the sensors in $x$ and $y$ would be perpendicular to the 1 g field, and the sensor in $z$ would be parallel to it. The devices in $x$ and $y$ would consequently experience different forces to the device in $z$.

\begin{figure}
\includegraphics[width=\columnwidth]{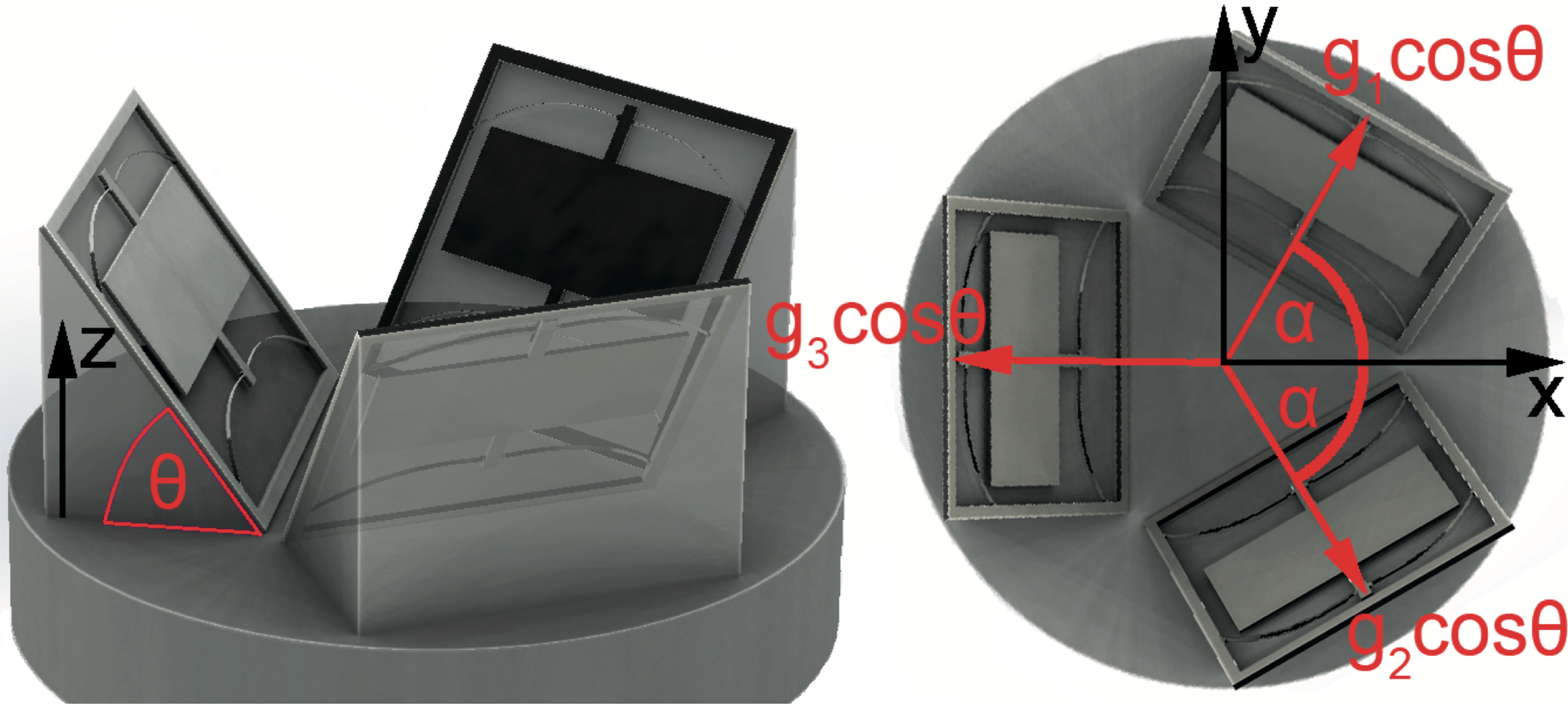}
\caption{\label{fig:Galperin}A Future Gravimeter Design. A computer generated image of a three-axis MEMS gravimeter in a Galperin configuration\,\cite{Galperin1955}.}
\vspace{-8pt}
\end{figure}

For a triaxial device such as that presented in figure \ref{fig:Galperin}, the acceleration in each of the three axes would be given by equations \ref{eq:Galperinz}, \ref{eq:Galperinx} and \ref{eq:Galperiny}:

\begin{equation}
g_z = \frac{(g_1 + g_2 + g_3)\sin{\theta}}{3}
\label{eq:Galperinz}
\end{equation}

\begin{equation}
g_x = \left((g_1 + g_2)\cos{\alpha}\right)\cos{\theta} - g_3\cos{\theta}
\label{eq:Galperinx}
\end{equation}

\begin{equation}
g_y = \left((g_1 - g_2)\sin{\alpha}\right)\cos{\theta}
\label{eq:Galperiny}
\end{equation}

To test the susceptibility of such a triaxial model to tilt, a geometry was tuned so that its minimum frequency would occur at the Galperin angle. Three identical models of this kind were then set up to match the configuration detailed in figure \ref{fig:Galperin}. The acceleration vectors of these models were then varied to simulate the effect of a tilting base plate. The displacement and resonant frequencies of each of the proof masses were measured, and a value of gravitational acceleration calculated using $g_i=\omega^2x$ (where $g_i$ is the output of one of the individual sensors). The total parasitic acceleration in $g_z$ (see eq. \ref{eq:Galperinz}) due to tilt was then plotted to ascertain whether tilt susceptibility was better in this configuration than for a single MEMS chip oriented vertically in a gravitational field. Figure \ref{fig:triax_plot} shows the results of this test. It can be seen that the triaxial configuration is less sensitive to tilt than a single vertical sensor. Since the triaxial system offers multi-axis sensitivity, as well as a reduced tilt susceptibility, it is clear that this design provide practical benefits in the field of gravimetry.


\begin{figure}
\includegraphics[width=\columnwidth]{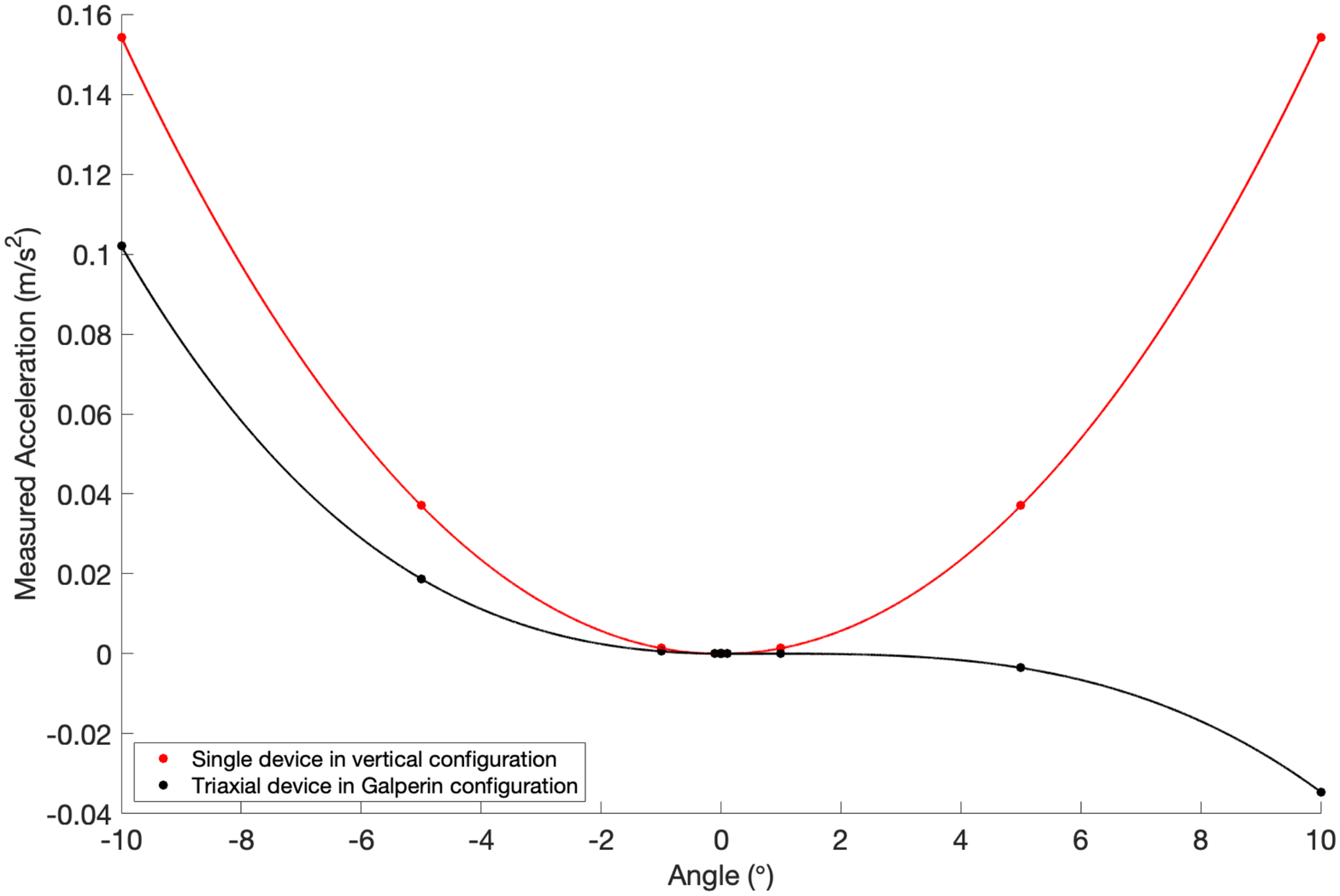}
\caption{\label{fig:triax_plot}Tilt Susceptibility Comparison. A comparison between the tilt susceptibility of a single vertical sensor, and that of three devices placed in a triaxial configuration.}
\end{figure}

\section{Methods}
\subsection{Fabrication of MEMS Gravimeters}
The fabrication of the MEMS devices used in these experiments were fabricated by Kelvin Nanotechnology. A similar fabrication process is discussed in detail in Middlemiss et al.\cite{Middlemiss2016}.

\section{Conclusion}
Here we taken the design of geometrical anti-springs from an analytical model, through a sequence of tests within a FE modelling environment to validate the power of this novel spring design. Although it is possible to construct analytical models of these complex non-linear systems, FE models offer superior power when designing systems for which multiple parameters need to be optimised. This is particularly the case when modelling the effect of the anisotropic crystalline structure of the fabrication material; something not possible with an equivalent analytical calculation. The ability to tune the frequency of these systems -- and in particular the angle at which the minimum in frequency occurs --  has enabled the design of a tri-axial MEMS gravimeter with acceleration sensitivity in all three axes. The FE simulations demonstrate that this configuration is less susceptible to tilt and from a practical gravimetry perspective it allows gravity measurements without the requirement to level the gravimeter before undertaking measurements which can reduce the time required to complete gravity surveys.

\clearpage
\section*{Acknowledgements}

The authors would like to thank Kelvin Nanotechnology who fabricated the final MEMS sensors discussed in this paper. They would also like to thank the staff and other users of the James Watt Nanofabrication Centre for help and support in undertaking the MEMS fabrication development.

This work was funded by the Royal Society Paul Instrument Fund (STFC grant number ST/M000427/1), and the UK National Quantum Technology Hub in Quantum Enhanced Imaging (EP/M01326X/1).

Dr R. P. Middlemiss is supported by the Royal Academy of Engineering under the Research Fellowship scheme (Project RF/201819/18/83).

\section*{Author Declaration}The authors declare no conflict of interest.

\section*{Author contributions}

\begin{itemize}
\item{Richard Middlemiss led the methodology of the etch process for the MEMS gravimeter. He conducted finite element modelling of the geometrical anti-spring. He led the writing of the manuscript.}

\item{Paul Campsie contributed to the initial design of the geometrical anti-spring flexures, and conducted preliminary finite element analysis.}

\item{William Cunningham provided consultancy on various technical aspects of the finite element modelling.}

\item{Rebecca Douglas contributed to the tuning of the geometrical anti-spring design.}

\item{Victoria McIvor contributed to the modelling of the effect of crystal orientation.}

\item{James Hough developed the methodology of utilizing geometric anti-springs for the MEMS gravimeter system and commented on the manuscript.}

\item{Sheila Rowan was responsible for the resources that were necessary to complete the project and commented on the manuscript.}

\item{Douglas Paul coordinated the fabrication of the MEMS gravimeter, and contributed to the manuscript preparation.}

\item{Abhinav Prasad conducted the validation of the tri-axial modelling, and contributed to the manuscript preparation.}

\item{Giles Hammond had the initial concept of the MEMS gravimeter, and had oversight of design process. He led the formulation of the analytical modelling, and conducted finite element modelling. He co-wrote parts of the manuscript with Richard Middlemiss.}

\end{itemize}

\section*{Additional Information}
\begin{itemize}
\item{The research data relevant to this letter are stored on the University of Glasgow's Enlighten Repository}
\item{Correspondence and requests for material should be addressed to richard.middlemiss@glasgow.ac.uk or giles.hammond@glasgow.ac.uk}
\end{itemize}


\begin{thebibliography}{10}
\expandafter\ifx\csname url\endcsname\relax
  \def\url#1{\texttt{#1}}\fi
\expandafter\ifx\csname urlprefix\endcsname\relax\def\urlprefix{URL }\fi
\providecommand{\bibinfo}[2]{#2}
\providecommand{\eprint}[2][]{\url{#2}}

\bibitem{Barnes2012}
\bibinfo{author}{Barnes, G.} \& \bibinfo{author}{Barraud, J.}
\newblock \bibinfo{title}{{Imaging geologic surfaces by inverting gravity
  gradient data with depth horizons}}.
\newblock \emph{\bibinfo{journal}{Geophysics}} \textbf{\bibinfo{volume}{77}},
  \bibinfo{pages}{G1--G11}
\newblock  (\bibinfo{year}{2012}).

\bibitem{Rim2015}
\bibinfo{author}{Rim, H.} \& \bibinfo{author}{Li, Y.}
\newblock \bibinfo{title}{{Advantages of borehole vector gravity in density
  imaging}}.
\newblock \emph{\bibinfo{journal}{Geophysics}} \textbf{\bibinfo{volume}{80}},
  \bibinfo{pages}{G1--G13}
\newblock  (\bibinfo{year}{2015}).

\bibitem{Kaufmann2014}
\bibinfo{author}{Kaufmann, G.}
\newblock \bibinfo{title}{{Geophysical mapping of solution and collapse
  sinkholes}}.
\newblock \emph{\bibinfo{journal}{Journal of Applied Geophysics}}
  \textbf{\bibinfo{volume}{111}}, \bibinfo{pages}{271--288}
\newblock  (\bibinfo{year}{2014}).

\bibitem{Butler1984}
\bibinfo{author}{Butler, D.}
\newblock \bibinfo{title}{{Microgravimetric and gravity gradient techniques for
  detection of subsurface cavities}}.
\newblock \emph{\bibinfo{journal}{Geophysics}} \textbf{\bibinfo{volume}{49}},
  \bibinfo{pages}{1084 -- 1096}
\newblock  (\bibinfo{year}{1984}).

\bibitem{Romaides2001}
\bibinfo{author}{Romaides, A.} \emph{et~al.}
\newblock \bibinfo{title}{{A comparison of gravimetric techniques for measuring
  subsurface void signals}}.
\newblock \emph{\bibinfo{journal}{Journal of Physics D: Applied Physics}}
  \textbf{\bibinfo{volume}{34}}, \bibinfo{pages}{433--443}
\newblock  (\bibinfo{year}{2001}).

\bibitem{Gasperikova2008}
\bibinfo{author}{Gasperikova, E.} \& \bibinfo{author}{Hoversten, G.~M.}
\newblock \bibinfo{title}{{Gravity monitoring of CO2 movement during
  sequestration: Model studies}}.
\newblock \emph{\bibinfo{journal}{Geophysics}} \textbf{\bibinfo{volume}{73}},
  \bibinfo{pages}{1ND--Z105}
\newblock  (\bibinfo{year}{2008}).

\bibitem{Nishijima2016}
\bibinfo{author}{Nishijima, J.} \emph{et~al.}
\newblock \bibinfo{title}{{Repeat Absolute and Relative Gravity Measurements
  for Geothermal Reservoir Monitoring in the Ogiri Geothermal Field, Southern
  Kyushu, Japan}}.
\newblock \emph{\bibinfo{journal}{IOP Conference Series: Earth and
  Environmental Science}} \textbf{\bibinfo{volume}{42}},
  \bibinfo{pages}{012004}
\newblock  (\bibinfo{year}{2016}).

\bibitem{Panisova2009}
\bibinfo{author}{Panisova, J.} \& \bibinfo{author}{Pasteka, R.}
\newblock \bibinfo{title}{{The use of microgravity technique in archaeology: A
  case study from the St. Nicolas Church in Pukanec, Slovakia}}.
\newblock \emph{\bibinfo{journal}{Contributions to Geophysics and Geodesy}}
  \textbf{\bibinfo{volume}{39}}, \bibinfo{pages}{237--254}
\newblock  (\bibinfo{year}{2009}).

\bibitem{Jin2013}
\bibinfo{author}{Jin, S.} \& \bibinfo{author}{Feng, G.}
\newblock \bibinfo{title}{{Large-scale variations of global groundwater from
  satellite gravimetry and hydrological models, 2002-2012}}.
\newblock \emph{\bibinfo{journal}{Global and Planetary Change}}
\newblock  (\bibinfo{year}{2013}).

\bibitem{Fores2017}
\bibinfo{author}{Fores, B.}, \bibinfo{author}{Champollion, C.},
  \bibinfo{author}{{Le Moigne}, N.}, \bibinfo{author}{Bayer, R.} \&
  \bibinfo{author}{Ch{\'{e}}ry, J.}
\newblock \bibinfo{title}{{Assessing the precision of the iGrav superconducting
  gravimeter for hydrological models and karstic hydrological process
  identification}}.
\newblock \emph{\bibinfo{journal}{Geophysical Journal International}}
  \textbf{\bibinfo{volume}{208}}, \bibinfo{pages}{269--280}
\newblock  (\bibinfo{year}{2017}).

\bibitem{Fernandez2017}
\bibinfo{author}{Fern{\'{a}}ndez, J.}, \bibinfo{author}{Pepe, A.},
  \bibinfo{author}{Poland, M.~P.} \& \bibinfo{author}{Sigmundsson, F.}
\newblock \bibinfo{title}{{Volcano Geodesy: Recent developments and future
  challenges}}.
\newblock \emph{\bibinfo{journal}{Journal of Volcanology and Geothermal
  Research}} \textbf{\bibinfo{volume}{344}}, \bibinfo{pages}{1--12}
\newblock  (\bibinfo{year}{2017}).

\bibitem{Carbone2017}
\bibinfo{author}{Carbone, D.}, \bibinfo{author}{Poland, M.~P.},
  \bibinfo{author}{Diament, M.} \& \bibinfo{author}{Greco, F.}
\newblock \bibinfo{title}{{The added value of time-variable microgravimetry to
  the understanding of how volcanoes work}}.
\newblock \emph{\bibinfo{journal}{Earth-Sci. Rev.}}
  \textbf{\bibinfo{volume}{169}}, \bibinfo{pages}{146--179}
\newblock  (\bibinfo{year}{2017}).

\bibitem{Aparicio2014}
\bibinfo{author}{Aparicio, S.-M.}, \bibinfo{author}{Sampedro, J.},
  \bibinfo{author}{Montesinos, F.} \& \bibinfo{author}{Molist, J.}
\newblock \bibinfo{title}{{Volcanic signatures in time gravity variations
  during the volcanic unrest on El Hierro (Canary Islands)}}.
\newblock \emph{\bibinfo{journal}{Journal of Geophysical Research: Solid
  Earth}} \textbf{\bibinfo{volume}{119}}, \bibinfo{pages}{5033--5051}
\newblock  (\bibinfo{year}{2014}).

\bibitem{Battaglia2008}
\bibinfo{author}{Battaglia, M.}, \bibinfo{author}{Gottsmann, J.},
  \bibinfo{author}{Carbone, D.} \& \bibinfo{author}{Fernandez, J.}
\newblock \bibinfo{title}{{4D volcano gravimetry}}.
\newblock \emph{\bibinfo{journal}{Geophysics}} \textbf{\bibinfo{volume}{73}},
  \bibinfo{pages}{WA3--WA18}
\newblock  (\bibinfo{year}{2008}).

\bibitem{Rymer2000a}
\bibinfo{author}{Rymer, H.}, \bibinfo{author}{Williams-jones, G.} \&
  \bibinfo{author}{Keynes, M.}
\newblock \bibinfo{title}{{Gravity and Deformation Measurements}}.
\newblock \emph{\bibinfo{journal}{Geophysical Research Letters}}
  \textbf{\bibinfo{volume}{27}}, \bibinfo{pages}{2389--2392}
\newblock  (\bibinfo{year}{2000}).

\bibitem{Middlemiss2016}
\bibinfo{author}{Middlemiss, R.~P.} \emph{et~al.}
\newblock \bibinfo{title}{{Measurement of the Earth Tides with a MEMS
  Gravimeter}}.
\newblock \emph{\bibinfo{journal}{Nature}} \textbf{\bibinfo{volume}{531}},
  \bibinfo{pages}{614--617}
\newblock  (\bibinfo{year}{2016}).

\bibitem{Campsie2016}
\bibinfo{author}{Campsie, P.}, \bibinfo{author}{Hammond, G.},
  \bibinfo{author}{Middlemiss, R.}, \bibinfo{author}{Paul, D.} \&
  \bibinfo{author}{Samarelli, A.}
\newblock \bibinfo{title}{{Measurement of Acceleration, US Patent Number:
  10802042}}
\newblock  (\bibinfo{year}{2015}).

\bibitem{Middlemiss2016a}
\bibinfo{author}{Middlemiss, R.~P.}
\newblock \emph{\bibinfo{title}{{A Practical MEMS Gravimeter}}}.
\newblock \bibinfo{type}{Ph\uppercase{D} \uppercase{T}hesis},
  \bibinfo{school}{University of Glasgow}
\newblock  (\bibinfo{year}{2016}).

\bibitem{Middlemiss2017}
\bibinfo{author}{Middlemiss, R.} \emph{et~al.}
\newblock \bibinfo{title}{{Field Tests of a Portable MEMS Gravimeter}}.
\newblock \emph{\bibinfo{journal}{Sensors}} \textbf{\bibinfo{volume}{17}},
  \bibinfo{pages}{2571}
\newblock  (\bibinfo{year}{2017}).

\bibitem{Bramsiepe2018}
\bibinfo{author}{Bramsiepe, S.~G.}, \bibinfo{author}{Loomes, D.},
  \bibinfo{author}{Middlemiss, R.~P.}, \bibinfo{author}{Paul, D.~J.} \&
  \bibinfo{author}{Hammond, G.~D.}
\newblock \bibinfo{title}{{A High Stability Optical Shadow Sensor with
  Applications for Precision Accelerometers}}.
\newblock \emph{\bibinfo{journal}{IEEE Sensors Journal}}
  \textbf{\bibinfo{volume}{18}}, \bibinfo{pages}{4108--4116}
\newblock  (\bibinfo{year}{2018}).
\newblock \eprint{1711.01253}.

\bibitem{Middlemiss2018}
\bibinfo{author}{Middlemiss, R.~P.} \emph{et~al.}
\newblock \bibinfo{title}{{Microelectromechanical system gravimeters as a new
  tool for gravity imaging}}.
\newblock \emph{\bibinfo{journal}{Phil. Trans. R. Soc. A}}
  \textbf{\bibinfo{volume}{376}}
\newblock  (\bibinfo{year}{2018}).

\bibitem{Carbone2020}
\bibinfo{author}{Carbone, D.} \emph{et~al.}
\newblock \bibinfo{title}{{The NEWTON-g Gravity Imager: Toward New Paradigms
  for Terrain Gravimetry}}.
\newblock \emph{\bibinfo{journal}{Frontiers in Earth Science}}
  \textbf{\bibinfo{volume}{8}}
\newblock  (\bibinfo{year}{2020}).

\bibitem{Ibrahim2008}
\bibinfo{author}{Ibrahim, R.}
\newblock \bibinfo{title}{{Recent advances in nonlinear passive vibration
  isolators}}.
\newblock \emph{\bibinfo{journal}{J.Sound Vib.}}
  \textbf{\bibinfo{volume}{314}}, \bibinfo{pages}{371--452}
\newblock  (\bibinfo{year}{2008}).

\bibitem{Bertolini1999}
\bibinfo{author}{Bertolini, A.}, \bibinfo{author}{Cella, G.},
  \bibinfo{author}{Desalvo, R.} \& \bibinfo{author}{Sannibale, V.}
\newblock \bibinfo{title}{{Seismic noise filters, vertical resonance frequency
  reduction with geometric anti-springs: a feasibility study}}.
\newblock \emph{\bibinfo{journal}{Nucl. Instr. Meth. Phys. Res. A}}
  \textbf{\bibinfo{volume}{435}}, \bibinfo{pages}{475--483}
\newblock  (\bibinfo{year}{1999}).

\bibitem{Cella2005}
\bibinfo{author}{Cella, G.}, \bibinfo{author}{Sannibale, V.},
  \bibinfo{author}{Desalvo, R.}, \bibinfo{author}{M{\'{a}}rka, S.} \&
  \bibinfo{author}{Takamori, A.}
\newblock \bibinfo{title}{{Monolithic geometric anti-spring blades}}.
\newblock \emph{\bibinfo{journal}{Nucl. Instr. Meth. Phys. Res. A}}
  \textbf{\bibinfo{volume}{540}}, \bibinfo{pages}{502--519}
\newblock  (\bibinfo{year}{2005}).

\bibitem{Acernese2015a}
\bibinfo{author}{Acernese, F.}, \bibinfo{author}{{De Rosa}, R.},
  \bibinfo{author}{Giordano, G.}, \bibinfo{author}{Romano, R.} \&
  \bibinfo{author}{Barone, F.}
\newblock \bibinfo{title}{{Low frequency inertial control strategy for seismic
  attenuation with multi-stage mechanical suspensions}}.
\newblock \emph{\bibinfo{journal}{Proc. SPIE}} \textbf{\bibinfo{volume}{9431}},
  \bibinfo{pages}{1--11}
\newblock  (\bibinfo{year}{2015}).

\bibitem{Abbott2016}
\bibinfo{author}{Abbott~et. al., B.~P.}
\newblock \bibinfo{title}{{Direct Observation of Gravitational Waves from a
  Binary Black Hole Merger}}.
\newblock \emph{\bibinfo{journal}{Phys. Rev. Lett.}}
  \textbf{\bibinfo{volume}{116}}, \bibinfo{pages}{1--16}
\newblock  (\bibinfo{year}{2016}).

\bibitem{Zhang2020}
\bibinfo{author}{Zhang, H.}, \bibinfo{author}{Wei, X.}, \bibinfo{author}{Gao,
  Y.} \& \bibinfo{author}{Cretu, E.}
\newblock \bibinfo{title}{{Frequency characteristics and thermal compensation
  of MEMS devices based on geometric anti-spring}}.
\newblock \emph{\bibinfo{journal}{Journal of Micromechanics and
  Microengineering}} \textbf{\bibinfo{volume}{30}}, \bibinfo{pages}{085014}
\newblock  (\bibinfo{year}{2020}).

\bibitem{Galperin1955}
\bibinfo{author}{Galperin, E.}
\newblock \bibinfo{title}{{Azimuthal method of seismic observations}}.
\newblock \emph{\bibinfo{journal}{Gostoptechizdat, Moscow}}
  \textbf{\bibinfo{volume}{80}}
\newblock  (\bibinfo{year}{1955}).

\bibitem{Shampine2000}
\bibinfo{author}{Shampine, L.~F.} \& \bibinfo{author}{Reichelt, M.~W.}
\newblock \bibinfo{title}{{Solving Boundary Value Problems for Ordinary
  Differential Equations in M atlab with bvp4c}}.
\newblock \emph{\bibinfo{journal}{MATLAB Technical Report}}
  \bibinfo{pages}{1--27}
\newblock  (\bibinfo{year}{2000}).

\bibitem{Hopcroft2010}
\bibinfo{author}{Hopcroft, M.}, \bibinfo{author}{Nix, W.} \&
  \bibinfo{author}{Kenny, T.}
\newblock \bibinfo{title}{{What is the Young's modulus of silicon?}}
\newblock \emph{\bibinfo{journal}{J. Microelectromech. Syst.}}
  \textbf{\bibinfo{volume}{19}}, \bibinfo{pages}{229--238}
\newblock  (\bibinfo{year}{2010}).

\bibitem{Cumming2014}
\bibinfo{author}{Cumming, A.} \emph{et~al.}
\newblock \bibinfo{title}{{Silicon mirror suspensions for gravitational wave
  detectors}}.
\newblock \emph{\bibinfo{journal}{Class. Quantum Grav.}}
  \textbf{\bibinfo{volume}{31}}, \bibinfo{pages}{025017}
\newblock  (\bibinfo{year}{2014}).

\bibitem{Wu2010}
\bibinfo{author}{Wu, B.}, \bibinfo{author}{Kumar, A.} \&
  \bibinfo{author}{Pamarthy, S.}
\newblock \bibinfo{title}{{High aspect ratio silicon etch: A review}}.
\newblock \emph{\bibinfo{journal}{Journal of Applied Physics}}
  \textbf{\bibinfo{volume}{108}}, \bibinfo{pages}{051101}
\newblock  (\bibinfo{year}{2010}).

\end{thebibliography}

\end{document}